\documentclass[journal]{IEEEtran}
\usepackage{amsmath}
\usepackage{cases}
\usepackage{amsthm}
\usepackage{amsfonts}
\usepackage{bbm}
\usepackage{amssymb}
\usepackage{calrsfs}
\usepackage{mathrsfs}
\usepackage{bm}
\usepackage{cite}
\usepackage{siunitx}
\usepackage{setspace}
\usepackage[caption=false,font=normalsize,labelfont=sf,textfont=sf]{subfig}
\usepackage{graphicx}
\DeclareMathAlphabet{\mathcal}{OMS}{cmsy}{m}{n}
\usepackage{color}
\usepackage{algorithm} 
\usepackage{algpseudocode} 
\begin{document}
\title{
Message Passing Based Demodulation of the Time-Encoded Digital Modulation Signal}
\author{Yuan Xu, and Wenhui Xiong
\thanks{
Yuan Xu and Wenhui Xiong are with National Key Laboratory of Wireless Communications, University of Electronic Science and Technology of China, Chengdu 611731, China (e-mail: 202511220630@std.uestc.edu.cn; whxiong@uestc.edu.cn).}
}
\markboth{ }%
{Shell \MakeLowercase{\textit{et al.}}: Message Passing-based fast demodulation of time encoded digital modulation signal}

\maketitle
 
\begin{abstract}
This letter proposes a  Message Passing (MP) based algorithm for demodulating the time-encoded digital modulation signal. The proposed algorithm processes the spikes generated by the Time-Encoding Machine (TEM) directly  on a per-spike basis, which enables  "on-the-fly" demodulatio with low latency.   
The computational complexity of our proposed method scales linearly  with the number of demodulated symbols, while the computational complexity of the  pseudo-inverse-based method scales cubically  with  the same number of demodulated symbols.  

\end{abstract}


\section{Introduction}
\label{sec1}

\IEEEPARstart{N}{on}-uniform sampling has been extensively studied by the signal processing society due to its capability to adapt to irregular signal patterns \cite{unser2000sampling,eldar2015sampling,renaudin2000asynchronous}. Recent years have seen growing interest in a specialized sampling scheme known as the Time-Encoding Machine (TEM) \cite{lazar2004perfect}, which performs signal sampling by generating the time sequence \( (t_k)_{k \in \mathbb{Z}} \) from the continuous input \( u(t) \). Unlike the conventional analog-to-digital converters (ADCs), which perform periodic sampling, the
TEM operates asynchronously.  This asynchronous operation eliminates the need for a high-precision clock, thereby reducing power consumption and minimizing electromagnetic interference \cite{naaman2022fri}. 
These advantages make TEM a compelling alternative sampling scheme for low-power applications.

\begin{figure}[ht]
  \centering
  \includegraphics[width=2.5in]{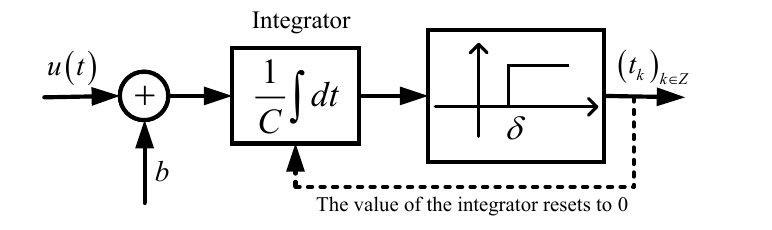}
  \caption{Integrate-and-fire  Time-Encoding Machine}
  \label{fig1}
\end{figure}

Fig.~\ref{fig1} shows the structure of integrate-and-fire TEM (IF-TEM) \cite{lazar2004perfect}. For the IF-TEM, a DC bias \( b \) is added to the input band-limited signal \( u(t) \) to ensure strict positivity; then the biased signal is scaled by a factor \( C \) and fed into the integrator. When the integrator reaches the predefined threshold \( \delta \), a spike is generated, and the integrator is reset. By repeating this process, the band-limited signal \( u(t) \) is encoded into a sequence of non-uniform, constant-amplitude spikes. 

Continuous signal recovery from spike sequences generated by TEM has been widely studied. Lazar \cite{lazar2004perfect} pioneered the field by proposing the Time Decoding Machine (TDM), a framework for reconstructing band-limited signals, and established the perfect recovery condition. Lazar models the TEM as a linear operation and uses the pseudo-inverse operation to recover the continuous signal.

While TDM operates in the time domain, Naaman \cite{naaman2022fri} studied the recovery method of the time encoded finite rate of innovation (FRI) signals, where the continuous signal is reconstructed in the frequency domain via the pseudo-inversion of the frequency domain sampling matrix. 

One of the limitations of the pseudo-inverse-based recovery method is its computational load due to the large matrix inversion. To address this limitation, Thao \cite{thao2023pseudo} proposed the projection onto convex sets (POCS) method, which replaces the pseudo-inversion of the linear sampling operator with iterative matrix multiplication for signal recovery.  The POCS method also enables minimum-norm reconstruction even when the perfect recovery conditions proposed by Lazar are not satisfied \cite{thao2023pseudo}.

Digital modulation signals, which convey information through discrete symbols, are widely used. The key focus of the receiver is not the reconstruction of the continuous signal, but the demodulation of the discrete symbols. 
To demodulate the discrete symbols, one can first recover the continuous signal, then use the classical matched filtering, followed by the hard decision.  Or,  one can follow the framework proposed by  \cite{lazar2004perfect} to compute the pseudo-inverse of the TEM, followed by a decision-making step for symbol demodulation.

The demodulation methods of time-encoded digital modulation signals, whether the reconstruction-then-demodulation approach or the pseudo-inverse-based method, share the same limitation of heavy computation.   
In particular, the pseudo-inverse-based method, despite its conceptual simplicity, needs large storage for buffering and high computational load of matrix inversion, both of which increase significantly as the signal duration grows \cite{thao2023pseudo}.  Similarly, while the POCS algorithm avoids the matrix inversion computation, it is inherently constrained by iterative operations that cause latency, making it unsuitable for time-sensitive applications.  

To address these limitations, we propose the Sliding Message Passing (SMP), which is based on the fact that any given modulation symbol only affects a limited number of spikes.  This resembles low-density parity-check (LDPC) codes, where each code symbol is governed  by a small number of parity-check equations. Inspired by this property, we propose the SMP algorithm, derived from the Message Passing (MP) framework, to enable low-latency demodulation.

In this letter, we first give a mathematical model for demodulating digital modulation signals in the IF-TEM-based receiver.  We explicitly map the spike sequence and digital modulation symbols to the node-edge structure of a factor graph for probabilistic inference. We then design the SMP algorithm by reformulating the MP algorithm into a per-spike form. The SMP algorithm enables incremental update of the symbol estimation by updating the Log Likelihood Ratio (LLR) of a given symbol, and the decision is made immediately upon receiving the spike at the symbol boundary.  This incremental update eliminates the need for the full spike sequence buffering.  This design not only reduces computational complexity and storage demands but also achieves "on-the-fly" demodulation, a critical feature for time-sensitive applications.

The remainder of this letter is organized as follows: Section II presents the system model for IF-TEM-sampled digital modulation signals, constructs the corresponding factor graph, and details the SMP algorithm. Section III analyzes  and compares the computational complexity and demodulation latency of SMP with  the mainstream demodulation methods. Section IV provides the simulation result, and compares the simulated BER results of different algorithms. Finally, Section V concludes the work.

\section{Sliding  Message Passing Algorithm}
This section details the  design of the SMP algorithm for demodulating digital modulation signals sampled by the IF-TEM. First, the block diagram of the IF-TEM-based digital communication system is presented, followed by the mathematical model of IF-TEM sampling. Subsequently, a factor graph is constructed to characterize the probabilistic relationships between the spike sequence and the modulation symbols. The MP-based algorithm is derived using LLR to lay the groundwork for the sliding optimization. Finally, the SMP algorithm is proposed to achieve low-latency "demodulate-while-receiving" processing, with its workflow elaborated.

\subsection{System Block Diagram and Signal Model}
\begin{figure}[ht]
  \centering
  \includegraphics[width=2.5in]{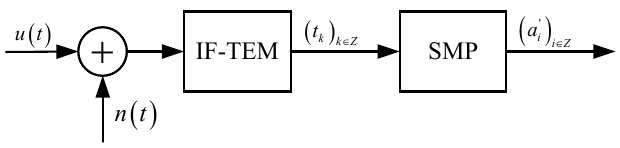}
 \caption{Block Diagram of the Demodulation Time-Encoded Signal Using the SMP Algorithm}
  \label{fig2}
\end{figure}

Fig.~\ref{fig2} illustrates the block diagram of the IF-TEM-based digital communication system proposed in this work, where the SMP module serves as the demodulation unit (detailed in subsection D). 

The baseband signal received by the system is given by 
\begin{equation}
\label{eq1}
    u(t) = \sum_{i \in \mathbb{Z}} \sqrt{E_s} a_i \, g(t - (i-1)T_s)
\end{equation}
where \( E_s \) is the energy per symbol, \( a_i\) denotes the digital modulation symbol, \( T_s = 1/R_s \) is the symbol duration, \(R_s\) represents the rate of the digital modulation symbol, and \( g(t) \) is the pulse shaping filter. For ease of presentation, we use Binary Phase Shift Keying (BPSK) as the exemplary modulation scheme, and the analysis can be extended to other modulation schemes.  For BPSK modulation,  \( a_i = \pm 1\)  with equal prior probabilities. In this work, we use the rectangular pulse shaping filter, i.e., \(g\left( t \right) = {{\mathbbm{1} _{\left[ {0,{T_s}} \right]}}\left( t \right)}/{\sqrt {{T_s}} }\). For any \( i \in \mathbb{Z} \), \( \|g(t - iT_s)\|_2 = 1 \) holds. We denote the space spanned by \( \{g(t - iT_s)\}_{i \in \mathbb{Z}} \) as \( \mathcal{B} \), which is a Hilbert space. Additionally, \(n\left( t \right)\) is additive white Gaussian noise (AWGN) that follows a Gaussian distribution \(\mathcal{N}\left( {0,{\sigma ^2}} \right)\).

When the signal with noise is input to the IF-TEM, a time sequence \({\left( {{t_k}} \right)_{k \in \mathbb{Z}}}\) is obtained. For the convenience of subsequent analysis, we define the spike-related vector \({\left( {{q_k}} \right)_{k \in \mathbb{Z}}}\) in the same manner as \cite{lazar2004perfect} for any \(k \in \mathbb{Z}\), i.e., 
\begin{equation}
\begin{aligned}
\label{eq2}
    {q_k}&\mathop= \limits^\Delta  \int_{{t_{k-1}}}^{{t_{k}}} {\left[ u\left( t \right) + n\left( t \right) \right]dt} \\
    &= \sqrt {{E_s}} \sum\limits_{i \in \mathbb{Z}} {{a_i}} \int_{{t_{k-1}}}^{{t_{k}}} {g\left( {t - (i-1){T_s}} \right)dt}  + \int_{{t_{k-1}}}^{{t_{k}}} {n\left( t \right)dt}\\
    &= C\delta  - bT_k.
\end{aligned}
\end{equation}

The vector \({\bf{q}} = {\left[ {{q_1},...,{q_k},...} \right]^\top}\) contains all the information of ${\left( {{t_k}} \right)_{k \in \mathbb{Z}}}$. It is easy to show that
\begin{equation}
\label{eq3}
    {\bf{q}} = \sqrt {{E_s}} {\bf{GA}} + {\bf{W}}
\end{equation}
where \({\left[ {\bf{G}} \right]_{k,i}}=g_{k,i} = \int_{{t_{k - 1}}}^{{t_k}} {g\left( {t - (i - 1){T_s}} \right)dt}\), \({\bf{A}} = {\left[ {{a_1},...,{a_i},...} \right]^T}\) is the transmitted symbol vector, \({\bf{W}} = {[{W_1},...{W_k},...]^\top}\) and it is easy to show that \({W_k}\) follows a Gaussian distribution \(\mathcal{N}\left( {0,{\sigma ^2}{T_k}} \right)\) with \({T_k} = {t_{k}} - {t_{k-1}}\).

\subsection{Factor Graph Construction}

\begin{figure}[ht]
  \centering
  \includegraphics[width=3.5in]{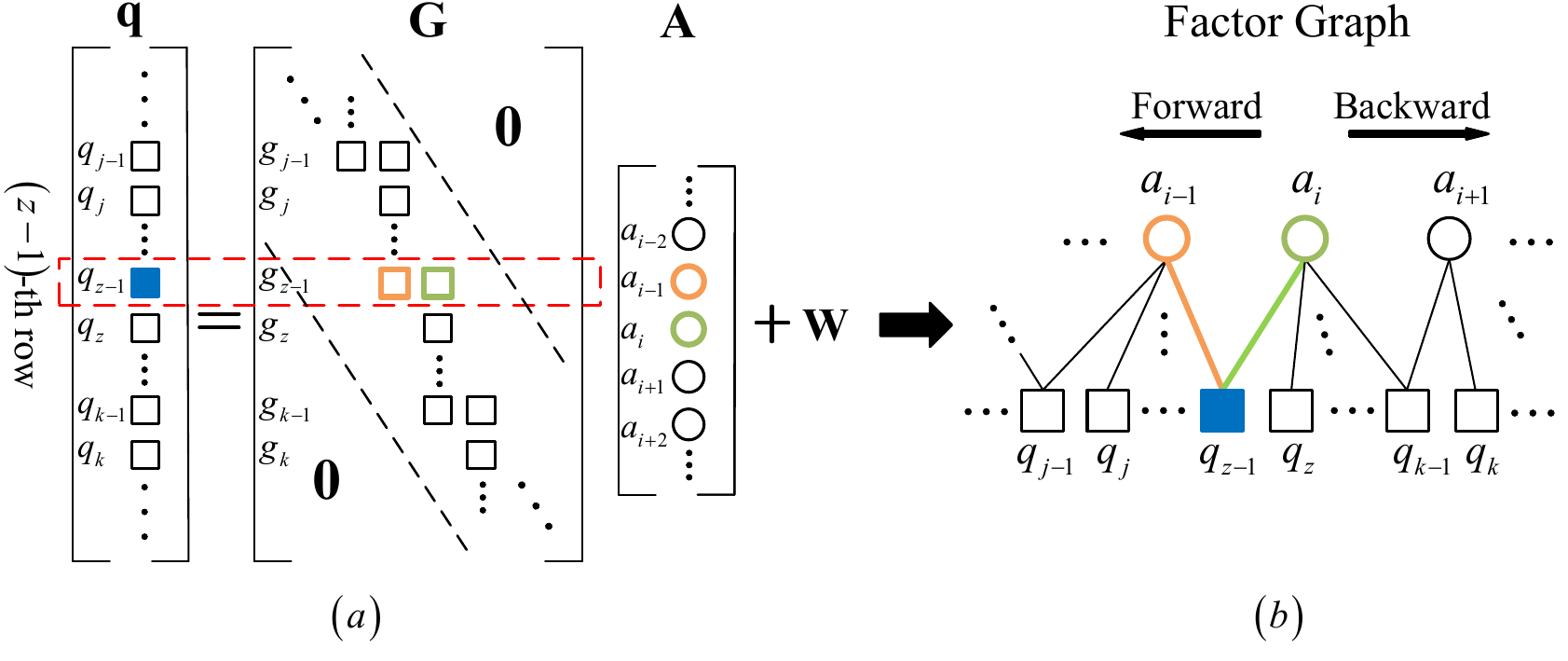}
  \caption{(a) IF-TEM sampling equation (3) (b)Factor graph with the variable node  \({a_i}\)  and  the factor node \({q_k}\)}
  \label{fig3}
\end{figure}

Fig.~\ref{fig3}(a) visualizes the TEM sampling equation (3), with specific emphasis on the sparse structure of the matrix \(\bf{G}\) . With the Gaussian noise  taken into account, the stochastic relationship between the TEM output vector \(\bf{q}\) and the modulation symbol vector \(\bf{A}\) can be expressed by the factor graph \cite{FactorGraph} shown in Fig.~\ref{fig3}(b), which models the probabilistic dependencies.

As illustrated in Fig.~\ref{fig3}(a), the matrix \(\bf{G}\) is a banded matrix with sparse non-zero elements. This sparsity arises from the time-localization of the BPSK symbols. Since the pulse shaping filter \(g(t)\) is non-zero only within \([0, T_s]\), the \(i\)-th symbol \(a_i\) only affects the spikes close to the interval \([(i-1)T_s, iT_s]\). As shown in Fig.~\ref{fig3}(a), each row of the matrix \(\mathbf{G}\) contains at most two non-zero columns, and these columns are adjacent.
In other words, as depicted in Fig.~\ref{fig3}(b), each variable node (the modulation symbol) is only connected to a finite number of factor nodes (the vector \(\bf{q}\)), and any two adjacent variable nodes share exactly one common factor node.

For instance, in Fig.~\ref{fig3}(a), in the \((z-1)\)-th row of the matrix \(\bf{G}\), only the elements in the \((i-1)\)-th and \(i\)-th columns, denoted as \(g_{z-1,i-1}\) and \(g_{z-1,i}\) are non-zero, while all other elements are zero. This means the \((z-1)\)-th  output of TEM, \({q_{z - 1}}\), is affected by only two symbols, \({a_i-1}\) and \({a_i}\). It can be expressed as
\begin{equation}
\label{eq5}
    \begin{aligned}
        {q_{z - 1}}
        &= \sqrt {{E_s}} {a_{i-1}}{g_{z-1,i-1}} + \sqrt {{E_s}} {a_i}{g_{z-1,i}} + {W_{z - 1}}.
    \end{aligned}
\end{equation}
\subsection{Message Passing-Based Demodulation Algorithm}
For the factor graph presented in Fig.~\ref{fig3}(b), the MP algorithm can be employed for demodulating \(\bf{A}\). In this section, we derive the demodulation algorithm based on an iterative LLR decoding algorithm \cite{moon2005error}.

To demodulate the $i$-th symbol under the condition that the vector \(\bf{q}\) is received, the LLR for  detecting the symbol $a_i$ is given by
\begin{equation}
\label{eq6}
    \lambda \left( {\left. {{a_i}} \right|{\bf{q}}} \right) = \log \frac{{p\left( {\left. {{a_i} = 1} \right|{\bf{q}}} \right)}}{{p\left( {\left. {{a_i} = - 1} \right|{\bf{q}}} \right)}} = \log \frac{{p\left( {\left. {\bf{q}} \right|{a_i} = 1} \right)}}{{p\left( {\left. {\bf{q}} \right|{a_i} = - 1} \right)}}.
\end{equation}
The second equality in (5) is because the transmitted symbols are equally probable.

Based on the structure of the matrix \(\bf{G}\) and the independence between noise and signal, (5) can be expressed as 
\begin{equation}
\label{eq7}
    \begin{aligned}
        \lambda \left( {\left. {{a_i}} \right|{\bf{q}}} \right) \!&=\!\lambda \left( {\left. {{a_i}} \right|{\{q_z,{z\in\mathcal{Q}_i\}}}} \right)\! +\!\lambda \left( {\left. {{a_i}} \right|{\{q_n,{n\notin\mathcal{Q}_i\}}}} \right)
    \end{aligned}
\end{equation}
where \({\mathcal{Q}_i}\) denotes the set of indices of factor nodes connected only to the single variable node \({a_i}\). The first term on the right-hand side of (6), calculated directly from the AWGN channel model, represents the intrinsic information for the detected symbol \({a_i}\); the second term, introduced through the connections between the adjacent symbols, represents the extrinsic information from other symbols.

For a  Gaussian channel, the intrinsic information can be computed as
\begin{equation}
\label{eq8}
    \lambda \left( {\left. {{a_i}} \right|{\{q_z,{z\in\mathcal{Q}_i\}}}} \right)= \sum\limits_{z \in {\mathcal{Q}_i}} {{L_s}{q_z}}
\end{equation}
where \(L_s = {2\sqrt {{E_s}} }/({{\sqrt {{T_s}} {\sigma ^2})}} \) is  the channel reliability \cite{moon2005error}.

As shown in Fig.~\ref{fig3}(b), the extrinsic information is introduced by the symbols adjacent to the detected symbol, which can be decomposed into the forward information and the backward information.  Take the symbol \({a_i}\) as an example,  the spike \({q_{z - 1}}\) represents the factor node connecting both the detected symbol \({a_i}\) and its forward neighbor \({a_{i - 1}}\), while the spike  \({q_{k - 1}}\) represents the factor node connecting both the detected symbol \({a_i}\) and its backward neighbor \({a_{i + 1}}\). By defining the set of indices of forward factor nodes as \(\mathcal{F}_i\) and the set of indices of backward factor nodes as \(\mathcal{B}_i\), the extrinsic information can be expressed as 
\begin{equation}
\label{eq9}
\begin{aligned}
        \lambda \! \left(\! {\left. {{a_i}} \right|\!{\{\!q_n,\!{n\!\notin\!\mathcal{Q}_i\!\}}}} \!\right)
       \! &=\!\lambda \! \left(\! {\left. {{a_i}} \right|\!{\{\!q_j,\!{j\!\in\!\mathcal{F}_i\!\}}}} \!\right)\! +\! \lambda \! \left( \!{\left. {{a_i}} \right|\!{\{\!q_k,\!{k\!\!\in\!\!\mathcal{B}_i\!\}}}}\! \right).
\end{aligned}
\end{equation}
We define  the forward information \(\eta _i\)  as 
\begin{equation}
\label{eq10}
\begin{aligned}
{\eta _i}&\mathop  = \limits^\Delta  \log \frac{{\sum\limits_{ \sim {a_i}} {p\left( {\left. {\left\{ {{q_j},j \in {\mathcal{F}_i}} \right\}} \right|\left\{ {{a_n},n \in \mathbb{Z}} \right\}} \right)} }}{{\sum\limits_{ \sim {a_i}} {p\left( {\left. {\left\{ {{q_j},j \in {\mathcal{F}_i}} \right\}} \right|\left\{ {{a_n},n \in \mathbb{Z}} \right\}} \right)} }}\\
 &= \log \frac{{\sum\limits_{{a_{i - 1}}} {p\left( {\left. {{q_{z - 1}}} \right|{a_i} = 1,{a_{i - 1}}} \right)  {\kappa_{{a_{i - 1}}}} } }}{{\sum\limits_{{a_{i - 1}}} {p\left( {\left. {{q_{z - 1}}} \right|{a_i} = - 1,{a_{i - 1}}} \right) {\kappa_{{a_{i - 1}}}} } }}.
\end{aligned}
\end{equation}
We define the function that depends only on the value of the symbol ${a_{i - 1}}$ as ${\kappa_{{a_{i - 1}}}}$, i.e.,
\begin{align}
   {\kappa_{{a_{i\!- \!1}}}}\!\mathop  = \limits^\Delta  \!\!\sum\limits_{ \sim {a_{i - 1}}}\!\! {p\!\left(\! {\left. {\left\{ {{q_j},j \!\in\! {F_i},j \!\ne\! z \!-\! 1} \right\}} \right|\!\left\{ {{a_n},n\! \in\! \mathbb{Z},n \!\ne\! i} \right\}} \right)}.
\end{align}

The equality in the second line of (9) arises from the facts that the elements of the spike vector \(\bf{q}\) are mutually independent given the symbol vector \(\bf{A}\), and for a factor node, its posterior probability depends only on the symbols it is connected to.

Since ${a_{i - 1}}$  only takes two values,  i.e., \(\pm 1\), we define  \(\gamma _{a_{i-1}} \mathop  = \limits^\Delta {\kappa_{{a_{i - 1}=1}}}/{{\kappa_{{a_{i - 1}=-1}}}}\). 
Thus, (9) can be simplified to 
\begin{equation}
\eta_i = \log \frac{
    \begin{aligned}
    &p_{i,i-1}^{z-1}(1,1)\gamma_{a_{i-1}} + p_{i,i-1}^{z-1}(1,-1)
    \end{aligned}
}{
    \begin{aligned}
    &p_{i,i-1}^{z-1}(-1,1)\gamma_{a_{i-1}} + p_{i,i-1}^{z-1}(-1,-1)
    \end{aligned}
}
\end{equation}
where \(p_{i,i-1}^{z-1}(a,b) \triangleq p\left(q_{z-1} \mid a_i=a, a_{i-1}=b\right)\) with \(a,b\in\{1,-1\}\) represents the likelihood function. It can be derived from the AWGN channel model as 
\begin{equation*}
    p_{i,i-1}^{z-1}(a,\!b)\!\propto \!\exp\! \left(\!\! - \frac{\left(q_{z-1}\!\!-\!\!\sqrt{E_s}\!\left(g_{z-1,i}a\!+\!g_{z-1,i-1}b\right)\!\right)\!^2}{{2{\sigma ^2}{T_{z - 1}}}} \!\!\right).
\end{equation*}

Similarly, the backward information in (8), denoted as $ \alpha_i$ 
\begin{equation}
\alpha_i \triangleq \log \frac{
    \begin{aligned}
    &p_{i,i+1}^{k-1}(1,1)\gamma_{a_{i+1}} + p_{i,i+1}^{k-1}(1,-1)
    \end{aligned}
}{
    \begin{aligned}
    &p_{i,i+1}^{k-1}(-1,1)\gamma_{a_{i+1}} + p_{i,i+1}^{k-1}(-1,-1)
    \end{aligned}
}.
\end{equation}

Based on the factor graph shown by Fig.~\ref{fig3}(b), we can derive  \(\eta _i^{\left( {l } \right)}\) and \(\alpha _i^{\left( {l } \right)}\) for the \(l\)-th iteration as
\begin{equation}
\begin{aligned}
\eta_i^{(l)}\!\!\! =\!\log \!\frac{
    p_{i,i-1}^{z-\!1}\!(1,\!1\!)e^{\lambda^{(l-1)}(a_{i-\!1}|\mathbf{q}) - \alpha_{i-1}^{(l-1)}} \!\!\!+ \!p_{i,i-1}^{z-\!1}\!(1,\!-\!1\!)
}{
    p_{i,i-\!1}^{z-\!1}\!(\!-\!1,\!1\!)e^{\lambda^{(l-\!1)}\!(a_{i-\!1}|\mathbf{q}) - \alpha_{i-1}^{\!(l-1)}} \!\!\!+\! p_{i,i-\!1}^{z-\!1}\!(\!-\!1,\!-\!1\!)
}
\end{aligned}
\end{equation}   
\begin{equation}
\begin{aligned}
\alpha_i^{(l)}\! \!\!=\!\log \! \frac{
    p_{i,i+1}^{k-\!1}\!(1,\!1\!)e^{\lambda^{(l-1)}(a_{i+\!1}|\mathbf{q}) - \eta_{i+1}^{(l-1)}}\!\!\! +\! p_{i,i+1}^{k-\!1}\!(1,\!-1\!)
}{
    p_{i,i+\!1}^{k-\!1}\!(\!-\!1,\!1\!)e^{\lambda^{(l-\!1)}\!(a_{i+\!1}|\mathbf{q}) - \eta_{i+1}^{\!(l-1)}}\!\!\! +\! p_{i,i+\!1}^{k-\!1}\!(\!-\!1,\!-\!1\!)
}
\end{aligned}
\end{equation}
According to (6), \(a_i\)'s LLR at \(l\)-th iteration is given by
   \begin{equation}
          \lambda^{(l)}\left( a_i \mid \mathbf{q} \right) = \sum_{z \in Q_i} L_s q_z + \eta_i^{(l)}  + \alpha_i^{(l)}.
   \end{equation}
The factor graph shown in Fig.~\ref{fig3}(b) has no cycles. In each iteration, the LLR of each symbol updates its forward and backward information  to incorporate information from one preceding bit and one subsequent bit, respectively. Consequently, for \(m\) symbols, the \(i\)-th symbol's LLR requires \(\max{(i-1,{m-i})}\) iterations to converge, i.e., all symbols’ LLRs converge after \(m-1\) iterations.
At this point, the Message Passing-based demodulation algorithm can be presented in Algorithm 1.
\begin{algorithm}[ht]
    \caption{Message Passing-based Demodulation Algorithm}
    \label{alg1}
    \begin{algorithmic}[1]
        \Require the pulse shaping filter \( g(t) \), the number of  symbols to be demodulated \(m\), the noise power \( \sigma^2 \), the sampling instants \( \{t_k\}_{k \in \mathbb{Z}} \), 
                 the IF-TEM parameters \( C, \delta, b \)
        
        \State Initialize \( \eta_i^{(0)} = 0 \) and \( \alpha_i^{(0)} = 0 \) for all \( i \)
        
        \For{ \( l = 0, 1, \dots, m-1 \) }
            \For{ \( i = 1, \dots, m \)}   
                \State Compute \(\lambda^{(l)}\left( a_i \mid \mathbf{q} \right)\) , \( \eta_i^{(l+1)} \),  and \( \alpha_i^{(l+1)} \)  by (15), (13), and (14)             
            \EndFor
        \EndFor
        
        \State \textbf{Decision}: For each \( i \), set \( \hat{a}_i = -1 \) if \( \lambda^{(m-1)}\left( a_i \mid \mathbf{q} \right) < 0 \), otherwise \( \hat{a}_i = 1 \)
    \end{algorithmic}
\end{algorithm}

\subsection{Sliding  Message Passing Algorithm}
\begin{figure}[ht]
    \centering
    \includegraphics[width=2.5in]{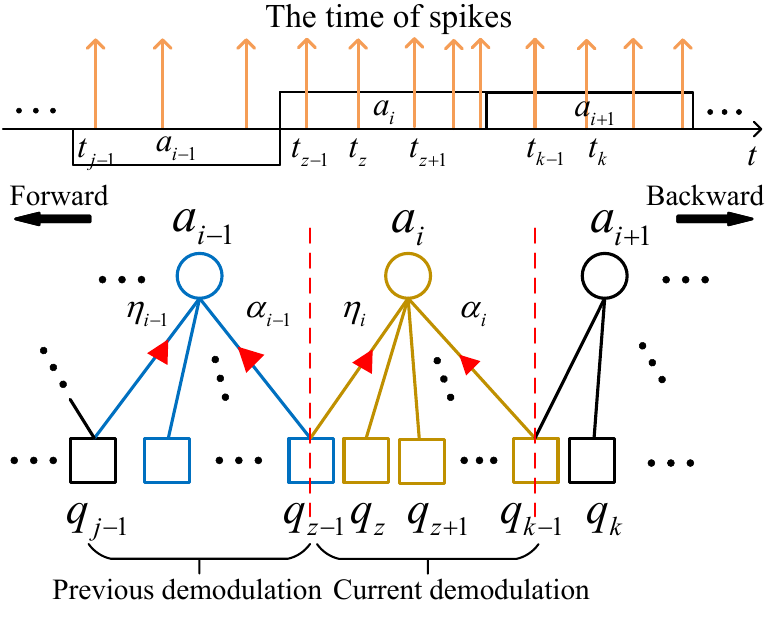}
    \caption{Schematic diagram of spike-by-spike information transmission}
    \label{fig4}
\end{figure}

The MP-based demodulation algorithm presented in the previous subsection requires receiving all the spikes of the entire signal duration before demodulation, which results in large latency. To reduce the latency of demodulation and enable demodulation while receiving, we propose the SMP algorithm, which updates the LLR of the symbol being demodulated when a new spike is generated by the TEM.

To better explain the SMP algorithm, we take the demodulation of the symbol $a_i$ as an example: Fig.~\ref{fig4} illustrates the messages from the factor nodes $q_{z-1},q_z, ... q_{k-1}$ to the variable node $a_i$. The demodulation of the symbol \(a_i\) relies on its LLR, which, as indicated in (6) and (8), consists of the intrinsic information, the forward information, and the backward information.  As shown in  Fig.~\ref{fig4}, when the spike at time \(t_{z-1}\) is obtained (indicated by the first red line in Fig.~\ref{fig4}), the forward information component of \(a_i\)'s LLR, i.e.,  $\lambda \left( {\left. {{a_i}} \right|{\{q_j,{j\in\mathcal{F}_i\}}}} \right)$ is updated based on the demodulation result of \(a_{i-1}\).  Thus, the forward information $\eta_i$ can be directly derived from (11)  as
\begin{equation}
\begin{aligned}
{\eta _i}\!& =\! \lambda \left( {\left. {{a_i}} \right|{q_{z-1},\hat{a}_{i-1}}} \right)=\log \frac{{p\left( {\left. {{q_{z - 1}}} \right|{a_i} = 1,{\hat{a}_{i - 1}}} \right)}}{{p\left( {\left. {{q_{z - 1}}} \right|{a_i} = - 1,{\hat{a}_{i - 1}} } \right)}}\\
=&\frac{{2\sqrt {\frac{{{E_s}}}{{{T_s}}}} \left( {{t_{z-1}}\! -\! i{T_s}} \right)\!\left(\! {q_{z-1} \!-\! {{\hat a}_{i - 1}}\sqrt {\frac{{{E_s}}}{{{T_s}}}} \!\left( {i{T_s} \!- \!{t_{z-2}}} \right)}\!\right)}}{{{\sigma ^2}{T_{z-1}}}}.
\end{aligned}
\end{equation}

As the process proceeds and new spikes are continuously generated by the TEM, the intrinsic information  of the symbol \(a_i\)'s LLR is updated incrementally. Specifically, for the spikes within \(a_i\)'s symbol duration \(\left[\left(i-1\right)T_b,iT_b\right]\), e.g., $q_z, q_{z+1},...,q_{k-2}$ the intrinsic information is updated using (7) on a spike basis. This process continues until the spike at time \(t_{k-1}\) is obtained. Note that \(t_{k-1}\)  is just greater than the symbol duration boundary of \(a_i\), and the backward information component of \(a_{i}\)‘s LLR \(\alpha_i\) can be expressed as

\begin{equation}
    \begin{aligned}
       & {\alpha _i} \! =\! \lambda \left( {\left. {{a_i}} \right|{q_{k-1}}} \right)\!=\!\log\! \frac{{p_{i,i+1}^{k-1}(1,\!1)\!+\! p_{i,i+1}^{k-1}(1,\!-1)}}{{p_{i,i+1}^{k-1}(-1,\!1)\!+\! p_{i,i+1}^{k-1}(-1,\!-1)}}
    \end{aligned}
\end{equation}

Thus, the LLR update for the symbol \(a_i\) is finalized and  \(\hat a_{i} \) is determined from the updated LLR. The SMP then slides to update the LLR of the next symbol \(a_{i+1}\) by treating \(\hat a_{i} \)  as the value of \(a_i\) for updating the forward information of \(a_{i+1}\).  
Thus, the update of the symbol \(a_i\)'s LLR can be done as
\begin{subnumcases}
{\lambda \left( {\left. {{a_i}} \right|{{\bf{q}}^{(k)}}} \right) = }
\lambda \left( {\left. {{a_i}} \right|{{\bf{q}}^{(k-1)}}} \right) + L_s q_k , & \text{if \(t_k < i T_s\)}\\
\lambda \left( {\left. {{a_i}} \right|{{\bf{q}}^{(k-1)}}} \right) + {\alpha _i} , &\text{else}.
\end{subnumcases}
The SMP algorithm is detailed in Algorithm 2.

\begin{algorithm}[ht]
    \caption{Sliding Message Passing (SMP) }
    \label{alg2}
    \begin{algorithmic}[1]
        \Require the pulse shaping filter $g\left( t \right)$, the noise power ${\sigma ^2}$, the sampling instants ${\left( {{t_k}} \right)_{k \in \mathbb{Z}}}$, and IF-TEM parameters $C,\delta ,b$
        
        \State Initialize \(\lambda \left( {\left. {{a_i}} \right|{{\bf{q}}^{(0)}}} \right) = 0\) for \(i = 1\) and initialize the index of the received spike as \(k=0\)
        
        \While{}
            \State Receive one spike, increment the count by 1, i.e., \(k=k+1\), and record the received time ${t_{k}}$.
            
            \If{${t_{k}} <  i {T_s}$} 
                Compute \(\lambda \left( {a_i} |{\bf{q}}^{(k)} \right) \) by (18a)
            \Else
                \State Compute \(\alpha _i\)  by (17) and  \(\lambda \left( {\left. {{a_i}} \right|{{\bf{q}}^{(k)}}} \right)\) by (18b)
                
                \State Make a decision: Set \({\hat a_i} = - 1\) if \(\lambda \left( {\left. {{a_i}} \right|{{\bf{q}}^{(k)}}} \right) < 0\), else set \({\hat a_i} = 1\)
                
                \State Set \(i = i + 1\)
                
                \State Compute \(\eta _i\) by (16) and set \(\lambda \left( {\left. {{a_i}} \right|{{\bf{q}}^{(k)}}} \right) = {\eta _i}\)
           
            \EndIf
        \EndWhile
    \end{algorithmic}
\end{algorithm}

\section{Complexity of the Proposed Algorithm}

In this section, we analyze the demodulation latency and complexity of the proposed SMP algorithm. The demodulation delay of the \(i\)-th symbol is defined as \(t_{\text{delay}} \!=\! \epsilon_i - iT_s\), where \(\epsilon_i\) denotes the time when the \(i\)-th symbol's demodulation is completed.

\textit{\textbf{Theorem 1}}:
Let \(u(t) = \sum_{i \in \mathbb{Z}} \sqrt{E_s} a_i \, g(t - (i-1)T_s)\) be a digital modulation signal, the demodulation delay of SMP for the TEM with an integrator constant \(C\) and a trigger parameter \(\delta\) is upper-bounded by \(t_{\text{delay}} \leq C\delta/{(b - c)}\).

\textit{\textbf{Proof}}:
As shown in Algorithm 2, SMP updates the LLR of one symbol at a time using simple division and addition. These operations  can be assumed to take less time than the spike interval ${T_{k}}= t_{k} - t_{k-1}$,  or equivalently, to be completed within the waiting period for the next spike.

In Step 6 of Algorithm 2,  when \(t_k > iT_s\) the LLR update completes, we can assume that the decision on the current symbol is made once the spike arrives, i.e., \(\epsilon_i \approx t_k\). Since \(t_{k-1} < iT_s < t_k\), \(t_{\text{delay}} = t_k - iT_s \leq T_k\). From \cite{lazar2004perfect}, \(T_k \leq C\delta/{(b - c)}\), which leads to \(t_{\text{delay}} \leq C\delta/{(b - c)}\).   

Next, we analyze the complexity of the SMP algorithm. Assume \(m\) symbols are to be demodulated, generating \(N\) spikes. As stated in \cite{lazar2004perfect}, the spike interval \(T_k \leq C\delta/{(b - c)}\).
For an observation time of \(mT_s\) (corresponding to \(m\) symbols), the number of spikes \(N \leq mT_s/({T_k)_\text{max}}\), which scales linearly with \(m\). The SMP algorithm performs only 1 or 2 computations per received spike;  its complexity is \(\mathcal{O}(m)\). For the pseudo-inverse-based method, the matrix \(\mathbf{G}\) shown in (3) has a dimension of \(N \times m\), and the complexity of matrix inversion is \(\mathcal{O}(Nm^2 + m^3) = \mathcal{O}(m^3)\). Therefore, the SMP algorithm has better complexity than the pseudo-inverse-based method.

\section{Simulation result and discussions}

In this section, we simulate the demodulation performance of the system given by  Fig.~\ref{fig2}.  In the simulation, we set the number of modulation symbols of each transmission to be  \(m=100\) with the symbol duration \(T_s = 1\ \text{ms}\).  
For the IF-TEM,  we set   $b = 3000$, $C = 1$, $\delta = 0.4$. We compare the BER performance of our proposed method against  conventional methods (e.g., pseudo-inverse-based demodulation) for different $E_s/N_0$.

The pseudo-inverse-based demodulation method can be performed in batch to obtain the least squares solution of the transmitted symbol vector \(\bf{A}\) from (3), i.e., 
\begin{equation}
  \bf{\hat{A} = \mathbf{sgn}(\bf{G}^{\dagger}\bf{q} )}
\end{equation}
where \( (\cdot)^{\dagger} \) denotes the Moore-Penrose pseudo-inverse operator, and \(\mathbf{sgn(\cdot)}\) is the element-wise signum function. The demodulation delay of this batch process is in the order of the duration of the  \(m\) symbols. To reduce the demodulation latency,  the pseudo-inverse-based method can be applied on a symbol-wise basis, i.e., the pseudo-inverse is computed for the subset of the matrix \(\bf{G}\) corresponding to the given symbol.  Specifically, as shown in (7), the subset \( \{q_k\}_{k \in  \mathcal{Q}_i} \) is within the duration of the symbol \( a_i \), and the demodulation of \( a_i \), denoted as \( \hat{a}_i \), can be derived by
\begin{equation}
\hat{a}_i = \mathbf{sgn}\left( \left( \mathbf{g}_i \big|{\mathcal{Q}_i} \right)^{\dagger} \left( \mathbf{q} \big|{\mathcal{Q}_i} \right) \right)
\end{equation}
where \(\left( \mathbf{g}_i \big|{\mathcal{Q}_i} \right)\) is the subvector of \(\bf{G}\)'s \(i\)-th column for \(k \in {\mathcal{Q}_i}\), and \(\left( \mathbf{q} \big|{\mathcal{Q}_i} \right)\) is the subvector of \(\mathbf{q}\) for \(k \in {\mathcal{Q}_i}\).

\begin{figure}[ht]
    \centering
    \includegraphics[width=2.5in]{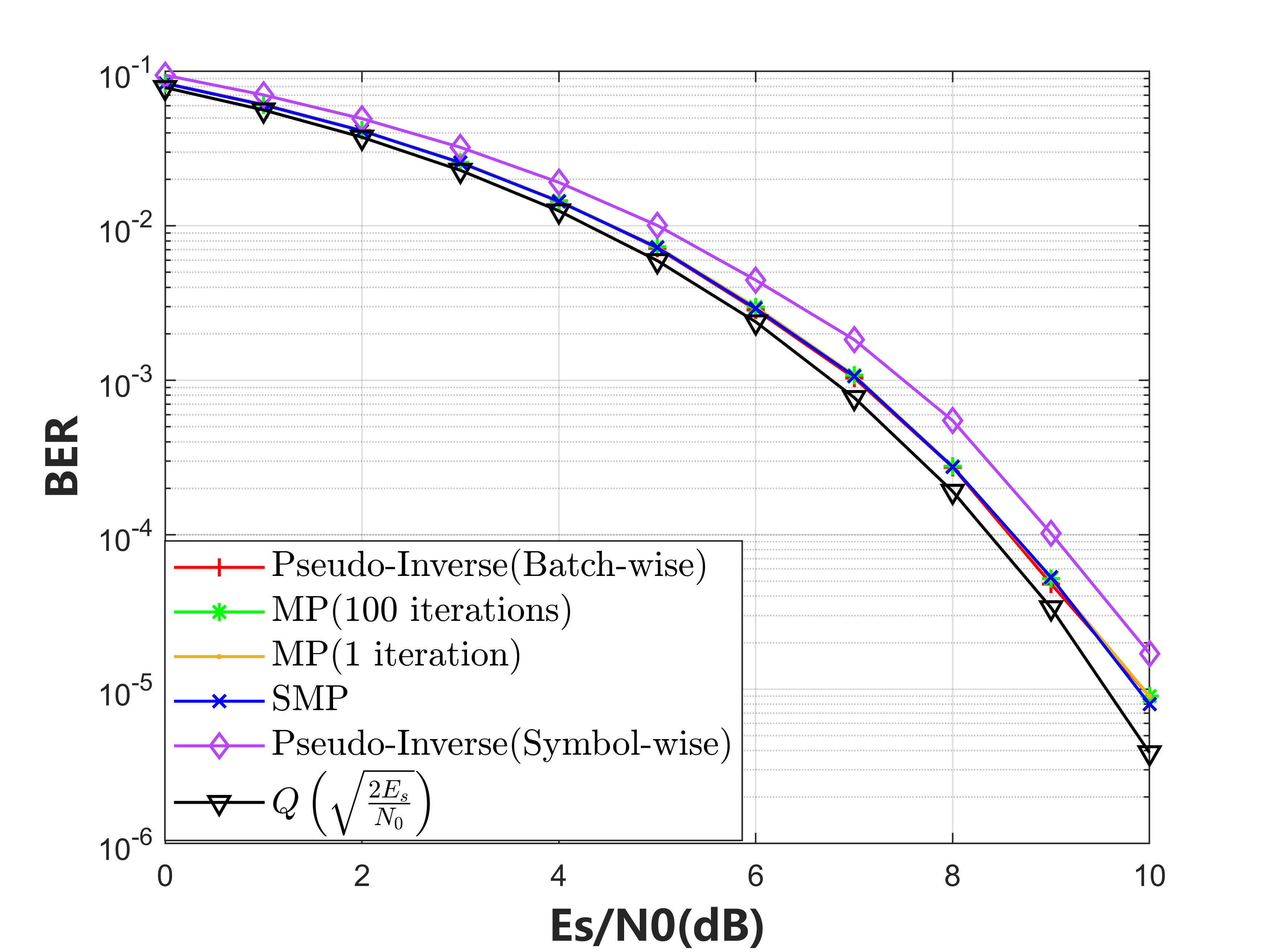} 
    \caption{Comparison of BER Performance of Different Signal Processing Methods at Varying \(E_s/N_0\)}
    \label{fig5}
\end{figure}

Fig.~\ref{fig5} compares the BER performance of different demodulation methods (Batch-wise Pseudo-Inverse, Symbol-wise Pseudo-Inverse, MP with 1/100 iterations, SMP) under AWGN with the theoretical BER of BPSK. 
As shown in Fig.~\ref{fig5}, the Batch-wise Pseudo-Inverse, MP(with similar performance for different numbers of iterations ), and SMP have similar BER performance, while the Symbol-wise Pseudo-Inverse is distinctly the worst performer.

For the MP-based method, the figure shows that one iteration is sufficient for good performance. 
This arises because each symbol is only dependent on its adjacent symbols (as depicted in the factor graph in Fig.~\ref{fig3}(b)). Thus, a single iteration of MP is enough to capture this dependency and reach optimal performance. The SMP is designed to transmit information from neighboring symbols, which approximates a single-iteration MP. Therefore, SMP achieves similar performance to the MP-based method. In contrast, the symbol-wise pseudo-inverse method, which processes each symbol independently, fails to use the inter-symbol dependency information, resulting in its poor BER performance.

Among these methods, the SMP enables reliable demodulation while ensuring low computational complexity and small demodulation latency.

\section{Conclusion}
This letter proposes a Sliding Message Passing (SMP), a fast demodulation algorithm for demodulating the time-encoded digital modulation signal.  The SMP enables low-latency "demodulation while receiving" processing.


\bibliographystyle{IEEEtran.bst}
\bibliography{IEEEabrv,mybib}


\end{document}